# Proposed Spreadsheet Transparency Definition and Measures

Craig Hatmaker
Harrisburg, PA
Craig_Hatmaker@Yahoo.Com

**ABSTRACT**

*Auditors demand financial models be transparent yet no consensus exists on what that means precisely. Without a clear modeling transparency definition we cannot know when our models are 'transparent'. The financial modeling community debates which methods are more or less transparent as though transparency is a quantifiable entity yet no measures exist. Without a transparency measure modelers cannot objectively evaluate methods and know which improves model transparency.*

*This paper proposes a definition for spreadsheet modeling transparency that is specific enough to create measures and automation tools for auditors to determine if a model meets transparency requirements. The definition also provides modelers the ability to objectively compare spreadsheet modeling methods to select which best meets their goals.*

## 1   INTRODUCTION – STATE OF THE ART

Transparency became a major topic after the accounting scandals of the 1990smoved the United States congress to introduce the "Corporate and Auditing Accountability, Responsibility, and Transparency Act of 2002" which later became known as the Sarbanes-Oxley Act of 2002, or SOX [Lasher 2008]. Major global financial modelling standards continue to maintain a focus on transparency. For example the T in FAST stands for transparent [FAST 2016].The words "transparency" and "transparent" appear numerous times on Corality's SMART webpage. Phrases like "improving the transparency" [*H.R.3763*] [FTC 2016], "enhance transparency" [FAST 2016]. "higher levels of transparency" [Schnackenberg, 2009] imply financial transparency is quantifiable yet none provide any transparency measures.

## 2   WHY MEASURE TRANSPARENCY

With transparency measures we can objectively identify:

- Opaque methods to exclude from modeling standards
- Opaque model sections to correct prior to model validation
- Models meeting transparency requirements and are thus ready for model validation
- Methods improving transparency to adopt to reduce model validation effort

Transparency measures open the possibility for automation tools to facilitate transparency audits.

## 3   TRANSPARENCY DEFINITION

To develop automation capable of quantifying transparency we must fully understand what Excel model transparency means. In looking at the plethora of researcher definitions, "*The common thread holding most definitions of transparency together is the notion that information must be disclosed to be transparent*" [Schnackenberg, 2009]

Merriam-Webster Dictionary also provides several transparency definitions including "easily understood". "Understandability" and "transparency" are linked with differences. To demonstrate a difference assume a model discloses all information and relationships used in a leveraged buyout model. According to Schnackenberg's "common thread" definition, the calculation is transparent but a toddler would have no understanding of how it works or what it means. Understandability is linked to reviewer abilities. This proposal removes the reviewer abilities variable by assuming all reviewers have requisite skills. For more on understandability in the Excel context see "Measuring Spreadsheet Formula Understandability" [Hermans 2012].

Two other Merriam-Webster definitions complement Schnackenberg's "common thread":

- Easily detected or seen through
- Characterized by visibility or accessibility of information

The key phrase "accessibility of information" is a good starting point. To clarify it we can borrow a tool from Information Engineering designed to turn high level, vague concepts into more actionable components: functional decomposition [Marin 1989][Sage 1991].

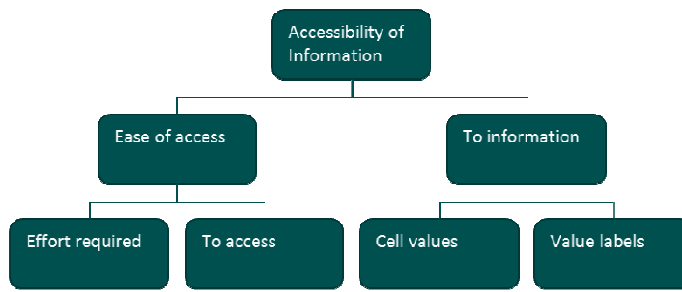

Let us start by decomposing "accessibility" to "ease of access". "Ease" can be further decomposed to "effort required," thus "accessibility" becomes "effort required to access."

To access something we must be able to detect it first and so "ease access" incorporates "easily detected" and what we are trying to detect and access is "information". "Information" in an Excel model context is cell values. Each cell displays a value. Each cell's value is a piece of information. The difference between data and information is data is a raw value and information adds meaning and context [Zins 2007] [Doyle 2014]. If a cell displays only raw data then for it to be information we must also find its label and that label must be sufficient to provide meaning and context. Thus, the result of our functional decomposition exercise is:

*Effort required to access cell values and value labels*.

That adequately defines cell surface level transparency but transparency as expressed in the definition phrase "seen through" infers a cell's sources, values from which a cell derives its value, must also be accessible. Thus, a proposed complete model transparency definition is:

*Effort required to access a cell's surface and source values and value labels.*

## 4   ESTABLISHING MEASURE

This paper proposes measuring effort. Merriam-Webster's thesaurus lists several synonyms and related words for "effort" which include "work" and "power". In physics power is the rate of doing work. In this case we need to quantify the rate of accessing cell values and labels.

### 4.1 Transparency Units of Measure

Excel provides several means by which we can find a cell's source values. They include:

- Highlight a reference and press F9 to see its value
- Select a named reference from the names drop-down to navigate to its location
- Click a formula in the formula bar to see all local reference locations
- Use menu option FORMULAS > Trace Precedents to point to local references.
- And more

Each of these methods is trivial by itself and so this paper proposes assigning each a work unit of one step where a 'step' involves something other than just looking such as mouse clicks or keyboard entries. Sometimes we must repeat or combine methods to display a cell's source values and labels. Each repetition is a step. Each additional method is a step. Thus, a cell's proposed transparency measure is the minimum number of steps required to access all of a cell's source values and labels. The proposed unit of measure is *steps from transparency*.

### 4.2 Transparency Terminology

To simplify discussions this paper proposes the following terminology:

**Labels**

This refers to that which provides meaning and context required to elevate raw data to information [Zins 2007] [Doyle 2014]. Labels can be placed in cells, data validation input messages, cell comments, documentation pages, external documentation or any other means accessible to inspectors and linked to values.

**Immediate Vicinity**

This indicates a group of cells can be displayed simultaneously within a monitor's window. This will vary by content, monitor resolution and use of "freeze panes."

**Surface Level Transparency**

This refers to values and labels viewable on the screen without having to select a cell.

**Source Level Transparency**

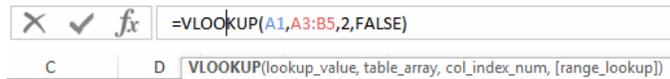

This refers to those references and literals from which the inspected cell derives its value. This only applies to cells with formulas. In the example above right we have selected a cell with a formula in which `A1` and `A3:B5` are source references, `2` and `FALSE` are source literals.

**Transparent**

A cell is transparent if no steps are required to see its surface values and labels as well as its source values and labels. Transparent is zero steps from transparency.

**Translucent**

A cell is translucent if any steps are required to find its values and labels as well as its source values and labels. Translucent is *n* steps from transparency.

**Opaque**

A cell is opaque if we cannot access either a cell's value, label, source values, or source labels.

**4.3 Convention**

To keep measurement quantities aligned with the concept of transparency this paper proposes expressing steps with negative values. Thus, the measure of transparency for any given model reference, function or formula is negative one times minimum steps required to find source values and labels. In this convention:

- Transparent: 0 steps from transparency is completely transparent
- Translucent: -# steps from transparency is less transparent
- Opaque: $-\infty$ steps indicates source values or sufficient labels are inaccessible

**5 LABELING REQUIREMENTS**

Occupied cells contain values and labels. Labels provide meaning and context for values. Label placement is subject to community standards while label content is subject to value type. Values can be categorized into the following types:

- Quantities
- Dates, Times and Durations

- Flags
- Identities
- Attributes

This paper proposes the following label content requirements based on value type.

### 5.1 Quantities

Quantities are magnitudes expressed as a number and reference. The reference includes a *type/kind* and *unit*[JCGM 2008]. In Excel terms, a cell containing a quantity value also requires labeling that includes *type/kind* and *Unit*.

| Subject | Examples |
| --- | --- |
| Quantities | Monies, periods, dimensions, etc. |
| Types/Kinds | Receivables, debt term, length, mass, area, volume, etc. |
| Units | USD, months, employees, kg, cm, g, etc. |

Units are also known as Units of Measure or UOM. Currency UOMs can be conveyed through formats that include currency symbols.

### 5.2 Dates, Times, and Durations

Dates and times mark when events occur or occurred. Durations are time quantities. Both are expressed in similar formats. These formats provide the UOM for various portions of the value. Because there are numerous formats their label must include their format as well as their subject.

| Subject | Examples |
| --- | --- |
| Dates and times | January 1$^{st}$, 2000; 12/1/2010; 11:00 AM; 01:23:14, etc. |
| Types/Kinds | Model Start, Debt Term, Period End, etc. |
| Formats | mmm dd yyyyy, mm/dd/yyyy hh:mm, hh:mm:ss, etc. |

### 5.3 Flags

Flags are Boolean values or switches. A flag's label must include the question the flag answers.

| Subject | Examples |
| --- | --- |
| Flags | True or False, 0 or 1, and Yes or No |
| Questions ('?' implied) | Due, Effective, Expired, etc. |

### 5.4 Identities

Identities are names, numbers, or codes that uniquely identify an entity. An identity label must include the subject.

| Subject | Examples |
|---|---|
| Identities | Employee IDs, Country names, Account Numbers, etc. |
| Types/Kinds | Employee, Customer, Account, etc. |

### 5.5 Attributes

Attributes are non-numeric object properties. An attribute's label must include its subject.

| Subject | Examples |
|---|---|
| Attributes | Variable Rate, Red, Round, Sour, Rough, etc. |
| Subjects | Loan Type, Color, Shape, Taste, Texture, etc. |

## 6 EXAMPLES

This paper provides the following examples to promote deeper understanding of what is meant by transparency in the Excel modeling context. These are only examples.

### 6.1 User Inputs/Assumptions

In the example at right users make entries in the "Value" column. User entries have no formulas; thus, only surface level transparency is evaluated. Each input is labeled with

| Label | Value | UOM/Format |
|---|---|---|
| Model Start | 1/1/2017 | mm/dd/yyyy |
| Model Duration | 3 | Months |
| Initial Investment | 100,000 | USD |

Type/Kind and Units/Formats within in the immediate vicinity. These are totally transparent.

### 6.2 Data Imports

In the example at right a dataset has been imported. The column heading provides Type/Kind labeling. Above quantity columns are units of measure/format labels. In

| | | mm/dd/yyyy | USD | USD |
|---|---|---|---|---|
| Account # | Account Name | Posted | Debit | Credit |
| 510 | Estimated Revenues 47,5 | 1/30/2017 | 8,838.00 | |
| 515 | Estimated Other Financing Sources | 1/25/2017 | | 4,767.00 |
| 540 | Appropriated Fund Balance | 1/19/2017 | 7,701.00 | |
| 900 | Appropriations | 1/25/2017 | 6,057.00 | |
| 905 | Other Financing Uses (Appropriations) | 1/12/2017 | | 4,179.00 |
| 890 | Unreserved, Undesignated Fund Balance 600 | 1/31/2017 | | 3,251.00 |

this example the surface level is totally transparent.

Data imported using MS Query exposes its source by right clicking in the data and selecting Table > External Data Properties > Connection Properties (icon) > Definition (tab) and examining "Command Text:" This could be considered one step and applying to the entire table.

Data imported using Power Query exposes its source by clicking in the data then right clicking on the data's query in the Workbook Queries panel and selecting Edit. This could be considered one step and applying to the entire table.

### 6.3 Literals in Formulas

We often find values expressed as constants embedded in formulas. In the example `=A11 * 12` there is a literal: `12`. We have no idea what 12 is other than a number. We do not know if it is a dozen, 12 inches in a foot, 12 months in a year, or something else completely. This source value is opaque.

### 6.4 Hidden Cells

A value that cannot be inspected is opaque. If a cell's value can be revealed by authorized inspectors using other means then those steps must be counted in the cell's transparency measure.

### 6.5 Error Cells

Cells displaying errors have inaccessible values and are opaque. Error values include:

- `#DIV/0!`
- `#N/A`
- `#REF!`
- `#NAME?`
- `#VALUE!`
- `#NUM!`

An exception is when errors are incorporated into downstream calculations as opposed to errors simply needing correction.

### 6.6 Unconstrained Indirect Reference

A cell's value that is derived from Excel's `INDIRECT()`, `OFFSET()`, `LOOKUP()`, `VLOOKUP()`, `HLOOKUP()`, or `INDEX()` functions that is not restricted to a specific cell range with appropriate labels is opaque because it is possible for the reference to point to cells with no value and/or label.

With care it is possible to make indirect reference functions translucent. The example at right provides model scenarios which can be selected via drop-down in `B20`. In cells `B21:B23` is this formula:

|   | A | B | C |
|---|---|---|---|
| 1 | | | |
| 2 | | | |
| 3 | Scenario #1 (tblScenario1) | | |
| 4 | | | |
| 5 | Label | Value | UOM/Format |
| 6 | Model Start | 1/1/2017 | mm/dd/yyyy |
| 7 | Model Duration | 3 | Months |
| 8 | Initial Investment | 100,000 | USD |
| 9 | | | |
| 10 | | | |
| 11 | Scenario #2 (tblScenario2) | | |
| 12 | | | |
| 13 | Label | Value | UOM/Format |
| 14 | Model Start | 6/1/2017 | mm/dd/yyyy |
| 15 | Model Duration | 3 | Months |
| 16 | Initial Investment | 75,000 | USD |
| 17 | | | |
| 18 | | | |
| 19 | Label | Value | UOM/Format |
| 20 | Scenario | tblScenario1 | |
| 21 | Model Start | tblScenario1 / tblScenario2 | m/dd/yyyy |
| 22 | Model Duration | 3 | Months |
| 23 | Initial Investment | 100,000 | USD |
| 24 | | | |

`=VLOOKUP([@Label],INDIRECT(OFFSET([#Headers],1,1, 1, 1)),2,FALSE)`

- `OFFSET()` is constrained to a single cell relative to the table's headers.
- `INDIRECT()` is constrained to a list of data validation values.
- `VLOOKUP()` is constrained to only what matches the current row's label in either of the scenario tables.

These constraints limit these indirect reference functions to cells with accessible values and labels.

### 6.7 Literal Constants in Functions

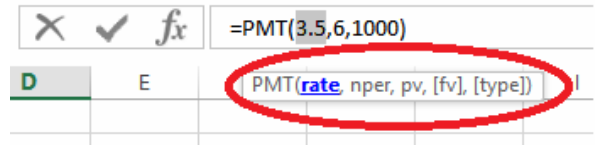

In the example at right `3.5` is a literal constant. Literal constants expose their values (`3.5`); thus, there are no steps required to find them. Finding their labels, which tell us what their value means, may require many steps.

When we use a literal constant in a function, the function may provide a parameter label sufficient for identifying what the literal is. To display the function's parameter labels we can double click the formula's cell or click the formula in the formula bar. Both methods expose a 'tooltip' below the cursor (red circle).

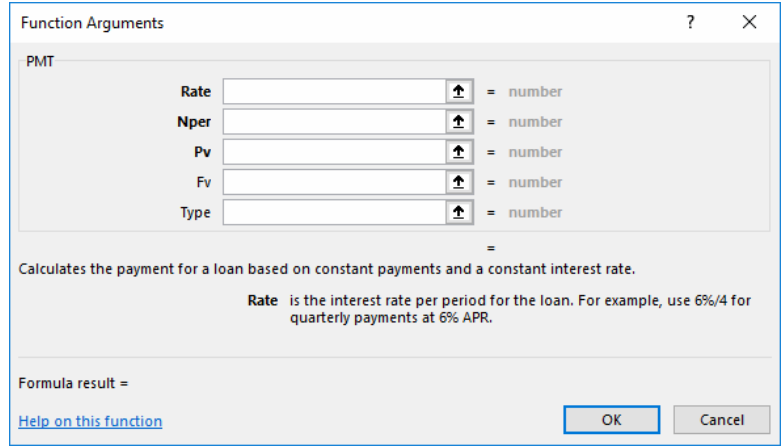

In this example we know `3.5` is a Rate. Rate is insufficient as it is lacking a subject (Interest) and a more meaningful unit of measure (APR). To find these required labels we must either double click PMT to bring up the function's help, or click the **fx** icon to display the *Function Arguments* dialog (shown right).

Sometimes parameter labeling is too vague. In the example we know `6` is the number of periods but we do not know if that is in weeks, months, quarters, etc.

Sometimes functions provide no meaningful labels at all. In such cases the literal is opaque.

A literal constant in a function's transparency is based on the function's parameter labeling being sufficient. For automation we can catalog Excel functions used and classify each parameter as sufficient or not. Thus, a literal constant in a function's transparency is -1 if the tooltip label is sufficient, -2 if we need to use the function's Help or Function Argument's dialog, or opaque if the function's labeling is insufficient.

**6.8 Cell/Range References**

At right is an example model section. Cell B1 is properly labeled; thus, its surface level transparency is 0 (completely transparent). If we select cell B1 its formula displays in Excel's formula bar.

Formulas require us to add B1's source level transparency by clicking anywhere in the formula bar. Excel color codes and highlights the cell references. Because everything is in the immediate vicinity we can see:

- The range `A3:B5` comprises a list of labeled values.
- Cell `A1`'s value is "`Initial Investment`" which is a label
- The literal "`2`" is identified by the formula tooltip as the column index number
- The flag "`FALSE`" is labeled "[range lookup]" which is inadequate labeling. Adequate labeling is found by clicking the **fx** icon and reading the Help text.

This example is -2 steps from transparency because a single click in the formula bar exposes all source values and labels (1$^{st}$ step) except the flag's label located in the help text (2$^{nd}$ step).

**6.9 Named Range References**

A named range is a defined name containing no functions or operators. It may contain a literal constant or cell/range reference. A name is a label and if it meets all labeling requirements eliminates the need to find a named range reference's label.

At right is an example using four named ranges in a formula placed in B1. This example is completely transparent because all surface values and labels as well as all source values and labels are visible simultaneously within the immediate vicinity.

If the named references were not in the immediate vicinity we could navigate to each named reference by using the Name drop-down list box located left of the formula bar, or F5, then selecting the name from the list. This adds one step to each reference in which case B1's transparency would by -4 steps from transparent (0 for B1's surface level transparency plus -1 step to find each of four named reference + 0 steps for each reference's surface level transparency)

**6.10 Structured References**

A structured reference is a type of dynamic named range generated automatically with tables. At right is a small table. A cell in the Net Income column is selected. It contains the formula:

   =[@EBIT]+[@Tax]

`[@EBIT]` is a structured reference that, like a cell reference, points to a location. While cell references point to locations in worksheets, structured references point to locations within tables. Cell references use worksheet names, column letters and row numbers as their name. Structured references use table names, column headings and special named regions as their name.

In this example, our selected cell has a transparency of 0 steps from transparent because `[@EBIT]` and `[@Tax]` are in the immediate vicinity along with required labeling.

**7   TIMING**

Transparency is only important when inspecting a model. Not all model users have the need, desire, time, skills, or authorization to appropriately inspect a model. If model information needs to be hidden for purposes of confidentiality or aesthetics the model is still 'transparent' if all pertinent source information is revealed when those qualified and authorized inspect it.

**8   IMPLEMENTING THIS NEW METRIC**

This paper proposes to implement this metric by assuming each cell is transparent and add to it associated component transparency measures. To describe the processes this paper uses pseudo code. Pseudo code provides human readable automation detail.

**8.1 Cell Surface Level Transparency**

Cell surface level transparency looks only at what is displayed in a worksheet cell. We can skip empty cells. Most model methodologies have regions set aside for labels which we can eliminate from scrutiny. Any cells displaying numeric values, regardless of location, must be measured.

```
If sufficient labeling not found then Transparency =-∞
    Else Transparency = Transparency – steps to find labeling
```

**8.2 Cell Source Level Transparency**

Cell source level transparency looks at what is inside a cell's formula, thus, this only applies to cells with formulas.

```
For each reference in a cell's formula
    If reference is a literal then
        If function's parameter labeling insufficient then Transparency =-∞
            Else Transparency = Transparency – 1
    Else
        Transparency = Transparency – steps to find reference cell
            +reference cell's surface level transparency
    End if
Next
```

### 8.3 Cell Transparency

Cell transparency is the sum of its surface and source levels. If any level is opaque, the entire cell is opaque.

```
Transparency = Cell Surface Level Transparency + Cell Source Level Transparency
```

### 8.4 Formula Transparency

A formula's transparency is the host cell's transparency.

### 8.5 Calculation Chain Transparency

A single result may include a set of formulas and references chained together. To calculate the chain's transparency we must total the transparencies of all cells in the chain. One way to do this is to start with the result and traverse the chain back until we end with cells without precedents.

```
For each reference in a cell's formula
    If reference is a literal then
        If function's parameter labeling insufficient then Transparency =-∞
            Else Transparency = Transparency – 1
    Else
        Transparency = Transparency + Reference's Chain Transparency
    End if
Next
```

### 8.6 Model Transparency

A model's transparency measure is the sum of all occupied cell transparency measures.

```
For each occupied cell in model
    If cell not a label then Transparency = Transparency + Cell's
        Transparency
Next
```

### 9   PRACTICAL APPLICATION

To improve model transparency we must first fix any opaque cells. After that we can consider (as one reviewer noted) if we have inadvertently made our model less transparent by replicating remote source values next to formula cells solely to improve each formula's transparency score. This just distributes the formula's transparency score to more cells with each cell worsening our overall model's transparency score. Our model's transparency score is an indicator of the total effort required to assess source values. Our goal should be to minimize such effort and thus make models as transparent as practical without increasing effort due to other measures such as complexity or readability.

## 10  WHAT WAS LEARNED

In developing this metric it became apparent that labeling is a crucial component with requirements varying by value type. Label placement impacts transparency and missing labels can make models opaque.

It also became apparent that some Excel functions can create opaque models giving legitimate reason for standards to bar them.

## 11  RECOMMENDATIONS

- Label all cell references appropriately.
- Label efficiently to reduce clutter.
- Use freeze panes to keep row and column labels in view (immediate vicinity).
- Seek alternatives to indirect references and if no practical alternatives exist make sure indirect references are constrained to occupied cells.
- When formulas require constants use appropriately named references instead of literals.
- When formulas require remote values use remote references instead of local cells daisy chained to remote cells.
- Use structured references when practical.
- Seek alternatives to error values if practical.
- Fix cells reporting errors not appropriately handled in downstream calculations.
- Favor model transparency over individual cell transparency.

## 12  SUMMARY

This proposal provides a means by which we can rationally measure model transparency and discern best practices through measurements rather than personal bias. We can also automate these methods to facilitate preparing models for audit.